# Fractal Analysis on Human Behaviors Dynamics

Chao Fan [1,*], Jin-Li Guo [1,†], Yi-Long Zha[2]

1. Business School, University of Shanghai for Science and Technology, Shanghai, 200093, PR China

2. Web Sciences Center, University of Electronic Science and Technology of China, Chengdu, 610054, PR China

**Abstract:** The study of human dynamics has attracted much interest from many fields recently. In this paper, the fractal characteristic of human behaviors is investigated from the perspective of time series constructed with the amount of library loans. The Hurst exponents and length of non-periodic cycles calculated through Rescaled Range Analysis indicate that the time series of human behaviors is fractal with long-range correlation. Then the time series are converted to complex networks by visibility graph algorithm. The topological properties of the networks, such as scale-free property, small-world effect and hierarchical structure imply that close relationships exist between the amounts of repetitious actions performed by people during certain periods of time, especially for some important days. Finally, the networks obtained are verified to be not fractal and self-similar using box-counting method. Our work implies the intrinsic regularity shown in human collective repetitious behaviors.

[*] Working organization: College of Arts and Science, Shanxi Agricultural University.

[†] Corresponding author E-mail addresses: Phd5816@163.com.

## 1. Introduction

In recent years, the studies of statistical characteristics in human behaviors attract much attention of researchers. In the past, it is generally assumed that human behaviors happen randomly, thus one of the conclusions of this assumption is that human behaviors can be described with Poisson process. Since 2005, as the inter-event time distributions of many human behaviors in daily life and work being investigated, such as E-mail and surface mail communications [1, 2], people found that these behaviors are totally different from the previous assumption: human behaviors exhibit the inhomogeneous feature with bursts and heavy tails, namely, the inter-event time distributions behave the right-skewed power-law shape. From then on, close attention have been paid to the research of human dynamics both in temporal scaling law such as web-browsing [3], short message communication [4], logistics operation [5] and spatial scaling characteristic in human mobility [6]. Moreover, many dynamic mechanisms have been proposed to explain the origin of the power-law distribution [1, 7, 8].

Given that people always repeat some actions in a certain period, the amount of events performed can be seen as a time series. Time series is defined as a set of quantitative observations recorded at a specific time and arranged in chronological order [9, 10]. It can be divided to two kinds, i.e. continuous ones and discrete ones, and generally time is considered as discrete variables in time series analysis which attracts special attention due to its practical and theoretical importance in physics, biology economics and society. Theoretical physics is one of the basic origins of the ideas and the methods. Applications of physical theories have led to fruitful achievements in this field. To cite an example, the complexity theory has played an important role in finding the non-trivial features in time series such as the long-range correlations, the scale invariance and so on. The studies of time series have great help to find the underlying rules of variables and forecast the future trend.

Complex network theory [11, 12] is a new branch in statistical physics, in which complex systems are described with networks. The nodes and edges represent the elements and the relationships between them respectively. The statistical characteristics of complex networks are mainly reflected by the measures like degree distribution, clustering coefficient and average shortest path length. Many networks in real world such as scientific collaboration network, Internet, air line network, protein interaction network, etc. show partially some of the following attributes like scale-free property, small-world effect, hierarchical structure as well as fractal and self-similar feature.

Recently, several efforts have been made to bridge time series and complex networks. Zhang et al. [13, 14] introduced the cycle networks for oscillation series, in which the segments between successive extremes are regarded as nodes and the distances in phase space between the nodes are used to construct the edges. Yang et al. [15] proposed a reliable procedure for constructing complex networks from the correlation matrix of a time series. Then Lacasa et al. [16] proposed the so-called visibility graph algorithm, in which the successive scalar time series points are mapped to nodes and each node is connected with all the other nodes covered within its visibility line. This new method rapidly attracted much attention for its simplicity and high efficiency, and a set of achievements have been obtained through it [17-22].

In our present research, the numbers of books borrowed by library readers in a certain period of time will be used to construct time series. In Sec.3, we verify that all the eight time series are long-term correlated with Rescaled Range Analysis by calculating the Hurst exponents and the length of non-periodic cycles. Then the time series are converted to networks through visibility algorithm in Sec.4, and those mapped networks exhibit scale-free, small-world and hierarchical features, confirming that the original time series have fractal feature. After that, in Sec.5, successive renormalization procedures which coarse-grain the networks into boxes containing nodes with a given size are performed to testify whether the visibility graphs are fractal and self-similar. To our surprise, all the networks are not fractal and self-similar. Finally, some conclusions and discussions are given in Sec.6.

## 2. Data specification

The data used in this paper are collected from the libraries of two universities located in different areas with distinct professional background. The data contain essential information to perform the statistical analysis such as users ID, exact time of borrowing and returning, etc. The sum of unique individuals in dataset A is 13,866, including undergraduates, postgraduates and teachers, together participating in a total of 139,606 transactions with time span from Sept.1st 2008 to Mar.30st 2009. As a comparison, all the 3,852 undergraduate users in dataset B come from the same grade and they created 328,795 items between Otc.12st 2005 and Jul.2nd 2009.

We take the amount of library loans, i.e. how many books have been borrowed, as the observation of time series using a unit in month, day and hour to study human behaviors. In addition, in view of inter-event time is a key measure in human dynamics, here we use time series as a tool to verify the correlation between numerical values in the inter-event time series. The basic statistical characteristics such as distribution of inter-event time of book borrowing or returning and the interval time between borrowing and returning are described in Ref.[23], and Table 1 gives the data volumes of the time series.

Table 1. The length of time series in our study‡

| Time series | Month | Day | Hour | Inter-event time |
|---|---|---|---|---|
| A | 7 | 180 | 1626 | 2000 |
| B | 46 | 1019 | 4000 | 2000 |

## 3. Fractal analysis of library loans time series

### 3.1 Hurst exponent and Rescaled Range Analysis

The Hurst exponent, namely, long-range correlation exponent is used as a measure of the long term memory of time series, i.e. the autocorrelation of the time series, and it has been extensively used in many fields such as the stock markets. In this section, we use Rescaled Range Analysis [24, 25]( $R/S$ for short in the following) to obtain the Hurst exponent of time series.

‡ Actually, there are much more amount of data in Inter-event time series of time series A and B, as well as in time series B with time unit in hours than the figures shown in Table 1. Due to that linear increase of data volume will bring geometric increase in computational complexity, only parts of the total data are considered. Meanwhile, non-workdays and nonworking time have been precluded.

Divide the time period into $A$ contiguous sub-periods of length $n$, such that $A \times n = N$. Each sub-period is labeled $I_a$, with $a = 1, 2, \cdots, A$ and then each element in $I_a$ is labeled $y_{k,a}$, with $k = 1, 2, \cdots, n$. For each sub-period $I_a$ of length $n$ the average is calculated as:

$$e_a = \frac{1}{n}\sum_{k=1}^{n} y_{k,a} \tag{1}$$

where $e_a$ is the average value of the items contained in sub-period $I_a$ of length $n$. Then we calculate the time series of accumulated departures $X_{k,a}$ from the mean at a given time $k$ for each sub-period $I_a$:

$$X_{k,a} = \sum_{i=1}^{k} (y_{i,a} - e_a), k = 1, 2, \cdots, n \tag{2}$$

Obviously, the last accumulated departure of every sub-period is 0.

The range that the time series covers from maximum values to minimum values within $I_a$ is defined as:

$$R_{I_a} = \max(X_{k,a}) - \min(X_{k,a}), 1 < k < n \tag{3}$$

where the value of the range depends on each given time. And the standard deviation for each sub-period $I_a$ can be expressed as:

$$S_{I_a} = \left[ \frac{1}{n}\sum_{k=1}^{n} (y_{k,a} - e_a)^2 \right]^{\frac{1}{2}} \tag{4}$$

Then the range for each sub-period $R_{I_a}$ is rescaled by the corresponding standard deviation $S_{I_a}$ depending on $A$ contiguous sub-periods of length $n$. Thus, the average value of $(R/S)_n$ for sub-period length $n$ is estimated by:

$$(R/S)_n = \frac{1}{A}\sum_{a=1}^{A} \left( \frac{R_{I_a}}{S_{I_a}} \right) \tag{5}$$

Repeat this process from Eq. (1) to Eq. (5) by increasing $n$ over successive integer values until $N/2$. The Hurst exponent may be estimated using least square regression of the following form:

$$\log(R/S)_n = H * \log n + \log c + \varepsilon_t \tag{6}$$

where $H$ is the Hurst exponent and can be estimated as the slope of the equation. Theoretically, the values of the Hurst exponent are in the range from 0 to 1 with a dividing point at 0.5, and the time series show distinct features upon the different value:

1) $H = 0.5$ means time series are standard Brownian motions and the event is independent, random and uncorrelated.
2) When $0.5 < H < 1$, the time series show positive correlation and persistence which is characterized by long-range memory effect. The persistence means that if the time series have been up or down, they are like to continue to be up or down in the future. Furthermore, there is long-term and non-periodical (namely definite periods don't exist) cycles in the system, which means it is difficult to forecast the future via current data. The limiting case $H = 1$ reflects a smooth signal.
3) Conversely, the time series will have anti-persistence and anti-correlation if $0 < H < 0.5$. That implies the time series more likely to move towards the opposite direction in the future compared with the past. The limiting case $H = 0$ corresponds to white noise, where fluctuations at all frequencies are equally present.

In a word, the Hurst exponent $H$ reflects the uneven level of time series. The closer $H$ gets to 0.5, the more noise and fluctuation there will be in the time series. Meanwhile, when $H$ deviates more from 0.5, the time series will be more regular and persistent.

### 3.2 Empirical analysis

We calculated the Hurst exponent of time series shown in Table 1, and the results are listed below:

Table 2. Results of $R/S$ analysis for library loans[§]

| Time series | Time unit | Hurst exponent | Length of non-periodical cycle |
|---|---|---|---|
| A | Day | 0.8217 | 10 |
| | Hour | 0.8050 | 89 |
| | Inter-event time | 0.6783 | 58 |
| B | Month | 0.7550 | 11 |
| | Day | 0.7470 | 233 |
| | Hour | 0.7267 | 1467 |
| | Inter-event time | 0.6893 | 32 |

Some conclusion can be attained as follows:

1) All the Hurst exponents surpass 0.5, thus the behaviors of book-borrowing don't obey random walk but exhibit long-term stability and correlation. The deviations of the library loans amounts tend to keep the same sign like the past.
2) With different time unit, both time series A and B have the similar value of $H$. It can be inferred that the extent of such correlation keep substantial agreement between different time scales. The fact that the exponent of A is a little bigger than that of B illuminates time series A has higher stability and persistence.
3) All the Hurst exponents get larger when the time units change from hour, day to month. Therefore the time series show much less noise and clearer tendency in bigger time scale. Obviously, this phenomenon is consistent with the actual situation that the total amounts of library loans are approximately consistent with each other in every month but fluctuate a lot in different hours even in days.
4) At the same time, the inter-event time series also have Hurst exponents bigger than 0.5. Combined with the fact that the inter-event times between two consecutive events of borrowing or returning follow power-law distribution, it can be concluded that human behaviors are uneven in general laws but regular in long-range correlation.

[§] Time series A with time unit month is excluded for its too few data volume.

Although the long-term memory processes can theoretically last forever, actually this effect might be finite. When long memory perishes, the trend of the time series will be changed at the crossover point. Technically, the point can be found in the $(R/S)_n \sim n$ double logarithmic plot where the slope changes for a certain value. Nevertheless, it is very difficult to identify that break point practically for the slope is so close before and after the crossover point. In order to find the exact point, we calculate the V-statistic which is defined as [25]:

$$V_n = \frac{(R/S)_n}{\sqrt{n}} . \tag{7}$$

The ratio of V-statistic is such that the rescaled range is scaled by the square root of time. More specifically, if the process is an independent, random process, the ratio will be constant and the slope of V-statistic plot is flat. On the other hand, if the process is persistent and the $(R/S)_n$ values are scaling at a faster rate than the square root time, and the slope of V-statistic plot is positively increasing, and vice versa. The crossover point $n_{\max}$ correspond to the length of non-periodic cycles, which means that the memory will be ineffective after $n_{\max}$ hours (or days, months) on average.

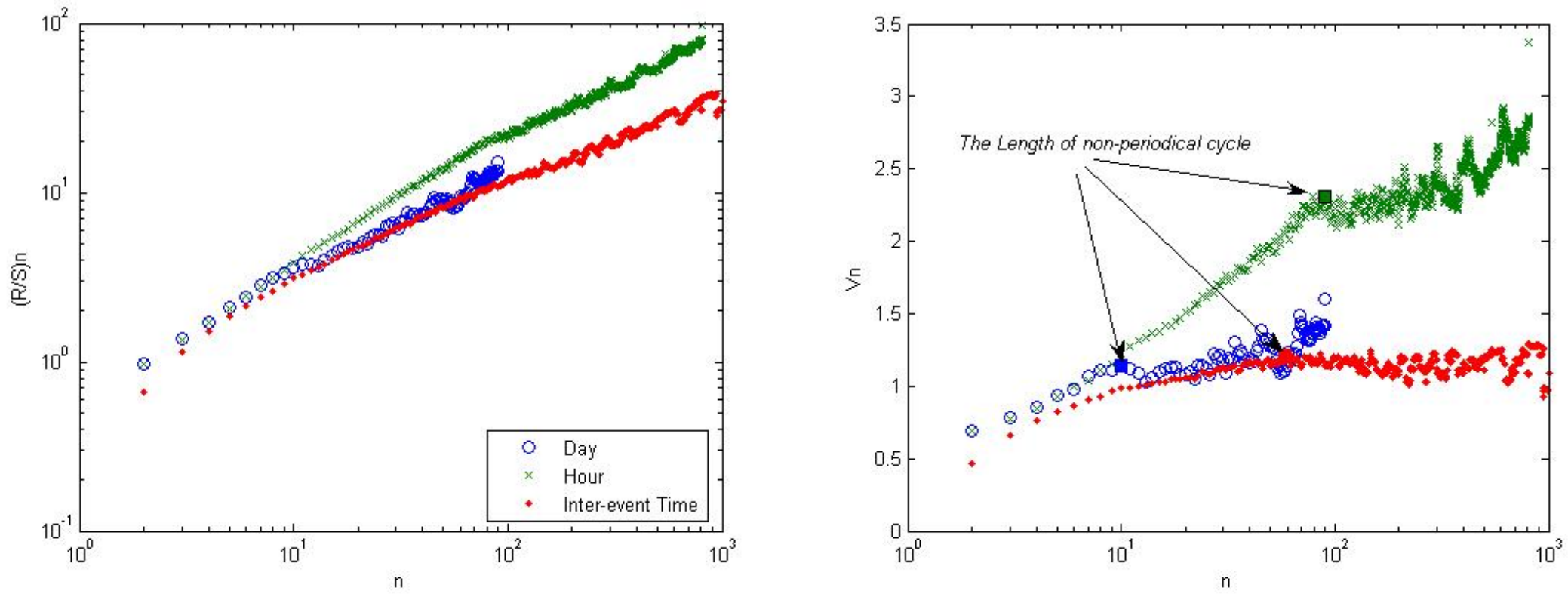

Figure 1. The plot of $(R/S)_n$ (left) and V-statistic (right) of time series A

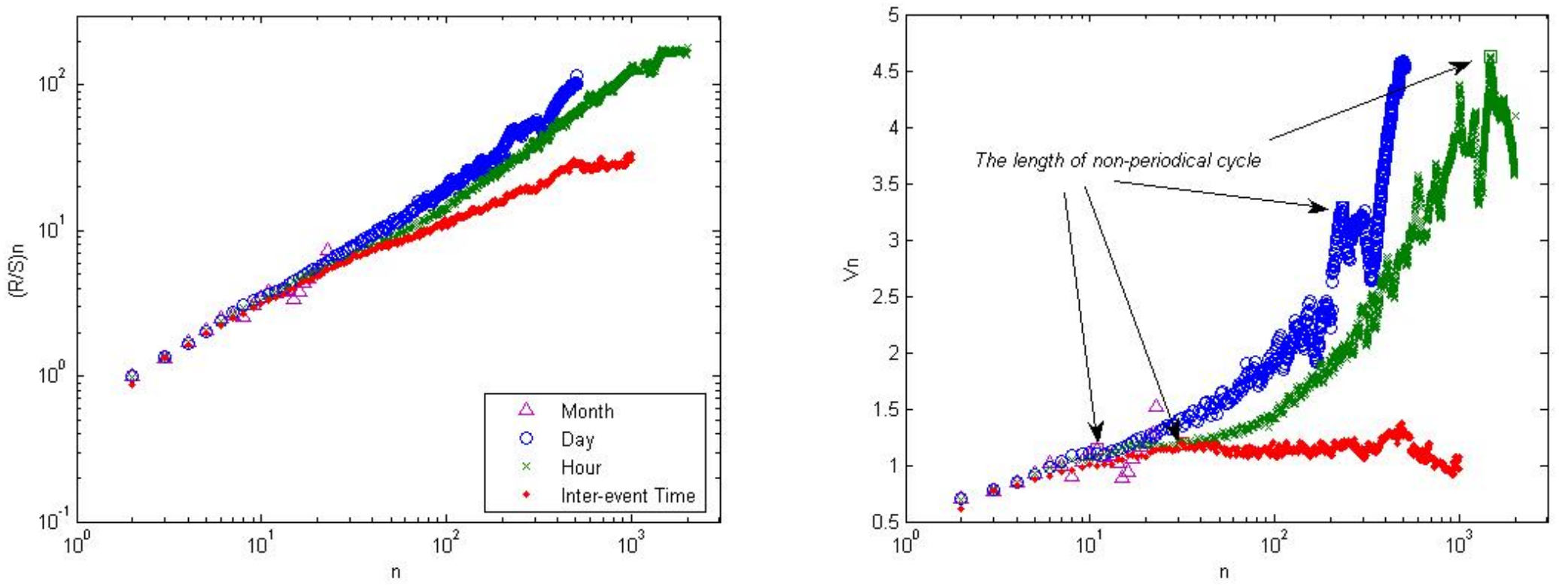

Figure 2. The plot of $(R/S)_n$ (left) and V-statistic (right) of time series B

We calculate the values of V-statistic following Eq. (7) and present the plots of $(R/S)_n$ and V-statistics against $\log(n)$ for the library loans in Fig.1 and Fig.2, upon which we obtain the length of non-periodic cycles shown in Table 2. That the slopes of V-statistic curves are increasing confirmed the memory effect in time series. Some features can be found in the results. Firstly, there exists obvious difference in the length of non-periodic cycles between time series A and B. It takes only 10 days to eliminate the memory effect in time series A, meanwhile as long as 11 months is needed in B. Secondly, taking consider the fact that the libraries are on service about 8.5 hours per day and 22.5 days per month, both two series exhibit consistency in the length of non-periodic cycles with different time scales. Therefore the cycling feature has no relation with time scale, which shows the fractal nature of the complex system. Another important phenomenon is such that the bigger $H$ is, the shorter the cycle length is. This is because the time series with higher Hurst exponent always have stronger long-term correlation between observations, thus events tend to be performed as the same pattern and then the cycle length will get shorter.

In a word, through analyzing the time series of library loans, fractal characteristic with long-term correlation can be obviously found in the book-borrowing behaviors and memory effect playing an important role in the actions. Furthermore, the existence of non-periodic cycles indicates that the human repetitious behaviors on collective level are not randomly happened but performed with some underlying regularity.

# 4. Statistical properties of complex network of human behaviors

As we introduced in Sec.1, among many ways to get time series converted to complex networks, visibility algorithm is a widely used one.

### 4.1 The principle of visibility algorithm

A visibility graph is obtained from the mapping of a time series into a network according to the following visibility criterion [16]: Two arbitrary data $(t_a, y_a)$ and $(t_b, y_b)$ in the time series have visibility and consequently become two connected nodes in the associated graph, if any other data $(t_c, y_c)$ such that $t_a < t_c < t_b$ fulfills:

$$y_c < y_a + (y_b - y_a)\frac{t_c - t_a}{t_c - t_a}. \tag{8}$$

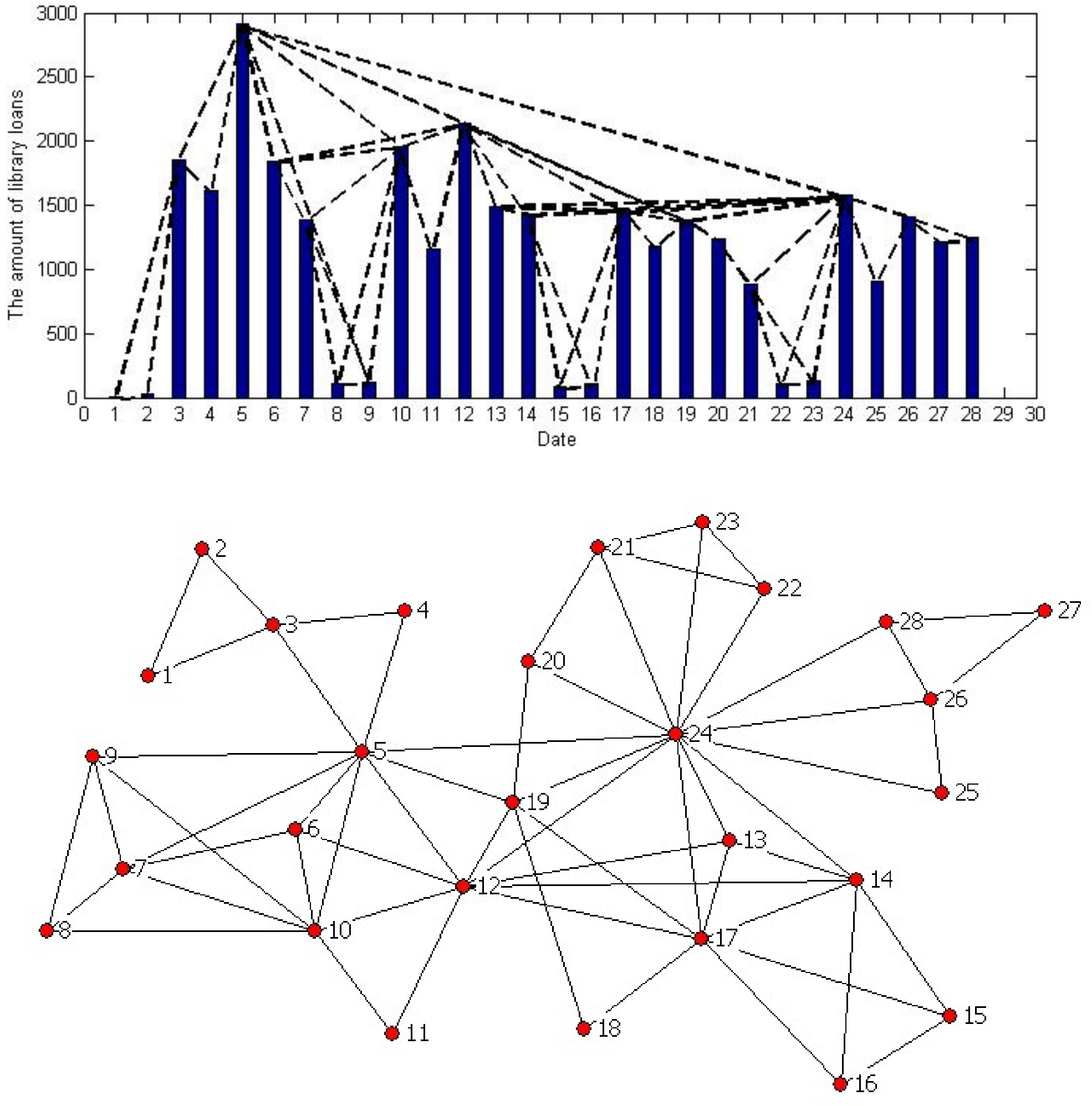

Figure 3: A typical example of visibility graph algorithm.

As shown in Fig.3, we take the daily amount of library loans from Oct.4th to 30st in dataset A as an example to illustrate this algorithm. In the upper panel, the data are displayed as vertical bars with heights indicating the values and the visibilities between data points are expressed as dashed lines. The converted network is shown in the lower panel, where the nodes correspond to series data in the same order and an edge connects two nodes if there is visibility between them.

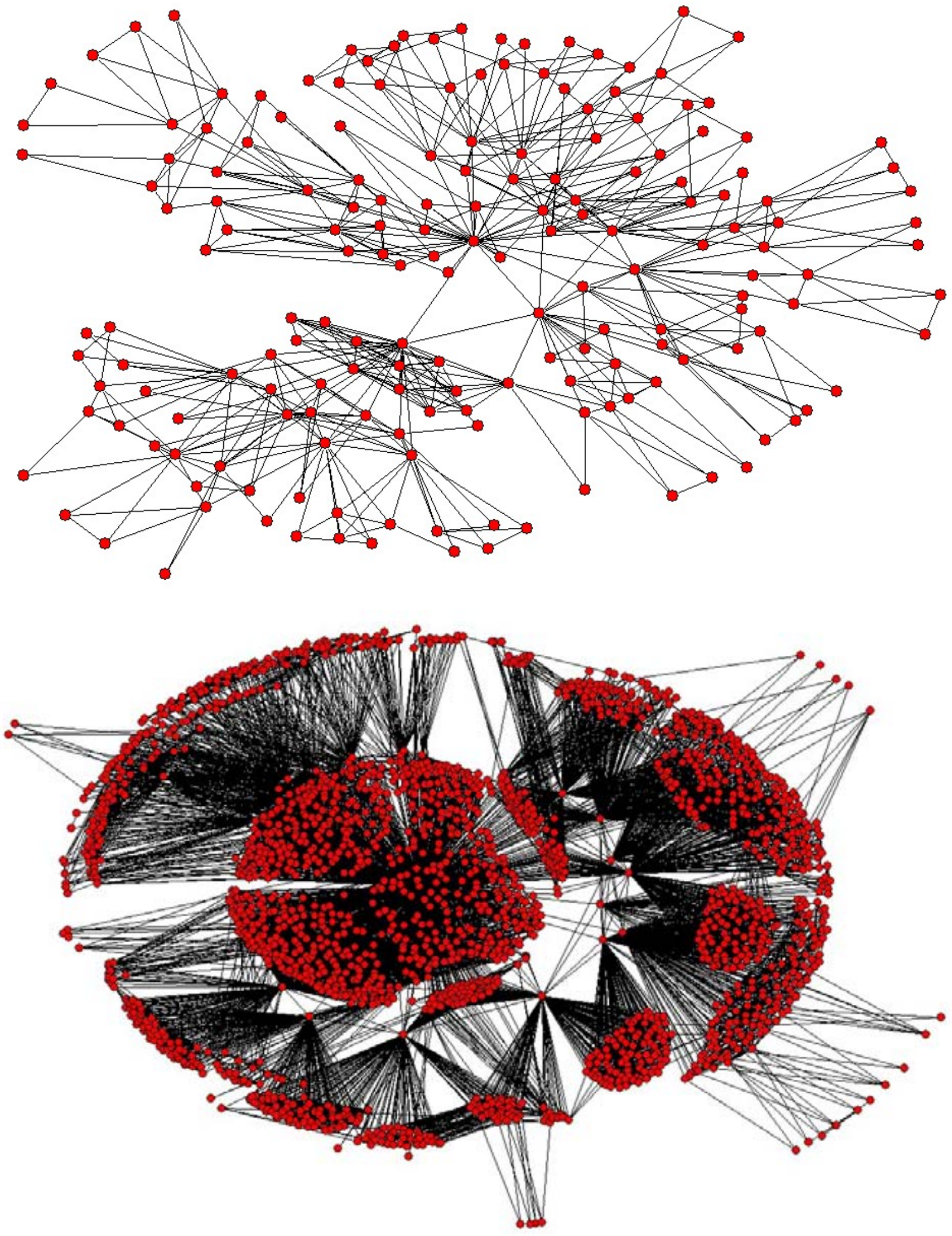

Figure 4. The network mapped from time series A with time unit in day (up) and inter-event time of time series B (down)

We can see from Fig.4 that the extracted graph is always connected, undirected, un-weighted and invariant under affine transformations of the series data. Further, it inherits several properties of time series in its structure. More specifically, periodic series convert into regular graphs, random series do so into exponential random graphs, and fractal series convert into scale-free networks, enhancing the fact that power-law degree distributions are related to fractality, something highly discussed recently.

**4.2 Empirical research**

All the 8 time series listed in Table 1 have been converted to visibility graphs using the algorithm introduced above. Then some parameters below are calculated:

1) Size of network $N$ : the total amount of nodes in the visibility graph.
2) Degree distribution $P(k)$ : the probability of a certain node to have degree $k$ . It is called scale-free network when degree distribution obeys a right-skewed power-law $P(k) \sim k^{-\gamma}$ .
3) Average degree of network $\langle k \rangle$ : the mean value of degrees of all the nodes in network, $\langle k \rangle = \sum_{i=1}^{N} k_i \Big/ N$ .
4) Average clustering coefficient $\langle C \rangle$ : the mean value of clustering coefficients of all the nodes in network, $\langle C \rangle = \sum_{i=1}^{N} C_i \Big/ N$ .
5) Diameter of network $D$ : the maximal distance between any pair of nodes, $D = \max_{i,j} d_{ij}$ , where $d_{ij}$ refers to the number of edges on the shortest path connecting node $i$ and $j$ .
6) Average path length $L$ : the mean value of the distance of any pair of nodes, $L = \frac{2}{N(N+1)} \sum_{i \geq j} d_{ij}$ .
7) The small-world effect: calculating the average path length step by step while increasing the number of nodes $N$ in network. If $L$ increases logarithmically along with $N$ , namely, $L(N) \sim \ln N$ , then the network is regarded with the feature of small world.

8) Hierarchical structure: weighted average values of clustering coefficients of nodes with degree $k$ were calculated as $\overline{C}(k) = \langle C \mid k \rangle = \sum_{i=1}^{n} C_i \Big/ n$, where $n$ refers to the number of different clustering coefficients a node with degree $k$ has. The network is considered to be hierarchically constructed if $\overline{C}(k) \sim k^{-\alpha}$.

All the empirical results of networks were shown in Table 3 and some conclusion can be deduced from the results as follows:

Firstly, all eight visibility graphs being scale-free networks verified that the time series of library loans to be with fractal characteristics. It also showed that the average degree, diameter and shortest path length all increase as the network grows. And the average clustering coefficient is getting larger at the same time due to the existence of hubs in scale-free networks enhance the clustering effect of the whole network.

Secondly, by comparing the power-law exponents $\gamma$ of visibility graphs and the Hurst exponents $H$ of time series, we can find inverse relationship between them. In other words, the bigger Hurst exponents correspond to the smaller power-law exponents, and that will bring higher regularity of time series. Recently, several researches [17, 18] present that for nonstationary time series there is a linear dependence between the values of $\gamma$ and $H$ like $\gamma = a - bH$. In our study, the relationships between them take the forms of $\gamma = 25.18 - 28.2H$ and $\gamma = 28.74 - 35.74H$ for datasets A and B respectively, which confirmed the conclusions in Ref.[17, 18] even though the values of $a$ and $b$ deviate numerically from them.

Thirdly, we can see from Fig.5 and Tab.3 obviously that all the networks mapped from time series have small-word phenomenon which means there are tight connections between nodes even they are located far away from each other for there are visibility lines between the corresponding data points in time series. Consequently, it can be deduced that some certain relations exist between the numbers of library loans in different time points. In another words, it is not unconnected between the past and the future in time series of human behaviors.

Table 3: Topological characteristics of the visibility graph

| Datasets | Time Unit | $N$ | $\gamma$ | $\langle k \rangle$ | $\langle C \rangle$ | $D$ | $L$ | $L(N) \sim N$ | $\overline{C}(k) \sim k$ | $r$ |
|---|---|---|---|---|---|---|---|---|---|---|
| A | Month | 7 | 1.2 | 2.571 | 0.614 | 3 | 1.714 | $L = 0.569\ln(N) + 0.632$ | $\overline{C} = 2.321k^{-1.30}$ | -0.4104 |
| | Day | 180 | 1.7 | 5.878 | 0.762 | 7 | 3.718 | $L = 0.619\ln(N) + 0.445$ | $\overline{C} = 3.074k^{-0.82}$ | 0.0996 |
| | Hour | 1626 | 2.5 | 8.576 | 0.773 | 10 | 4.958 | $L = 0.551\ln(N) + 0.779$ | $\overline{C} = 6.259k^{-0.93}$ | 0.0707 |
| | Inter-event time | 2000 | 3.8 | 9.804 | 0.838 | 6 | 2.819 | $L = 0.184\ln(N) + 1.424$ | $\overline{C} = 7.151k^{-0.96}$ | -0.1912 |
| B | Month | 46 | 1.6 | 4.609 | 0.704 | 7 | 2.994 | $L = 0.613\ln(N) + 0.620$ | $\overline{C} = 2.779k^{-0.90}$ | -0.0163 |
| | Day | 1019 | 2.3 | 7.030 | 0.763 | 10 | 4.660 | $L = 0.576\ln(N) + 0.812$ | $\overline{C} = 4.868k^{-0.92}$ | 0.0831 |
| | Hour | 4000 | 2.4 | 6.648 | 0.770 | 12 | 5.500 | $L = 0.569\ln(N) + 0.632$ | $\overline{C} = 3.347k^{-0.83}$ | 0.1492 |
| | Inter-event time | 2000 | 2.8 | 9.662 | 0.824 | 6 | 2.856 | $L = 0.211\ln(N) + 1.122$ | $\overline{C} = 7.482k^{-0.98}$ | -0.1147 |

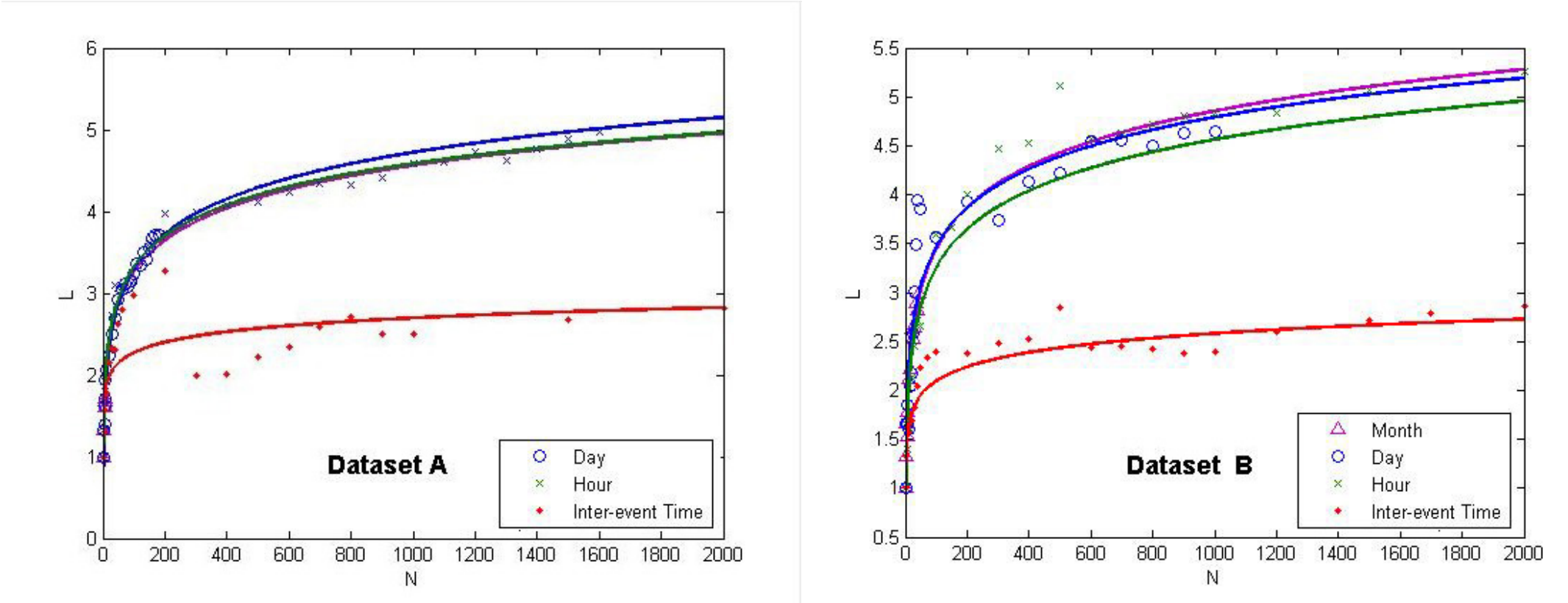

Figure 5. Plots of $L(N) \sim N$ to investigate the small world effect of networks.

Finally, the clustering coefficient decreases with degree by power-law testifies that the visibility graphs are hierarchically constructed. It has been proved that small number of hub-nodes own great number of links leaving most nodes with few edges connected may lead to scale-free property. Generally, nodes with low degree and high clustering coefficient are likely to form the densely connected sub-groups, which are connected with each other by the nodes with large degrees and small clustering coefficients. Then these sub-groups may connect with each other to form groups. Finally, these modular structures lead to the hierarchy of networks. The high degree of connection in hub-nodes in network denotes that some intrinsic relations exist between extreme points which not only have large values but also surrounded by small values in time series. Furthermore, that the values of $\alpha$ being slightly smaller than 1 (although with one exception) implies the networks are not strictly hierarchical. This can be seen in Fig.4 and Fig.6.

In a word, the visibility graphs converted from time series are scale-free, small-world and hierarchically constructed. Thus the original time series are deemed to be with fractal feature and there are close relationships between the data points, especially those extreme points.

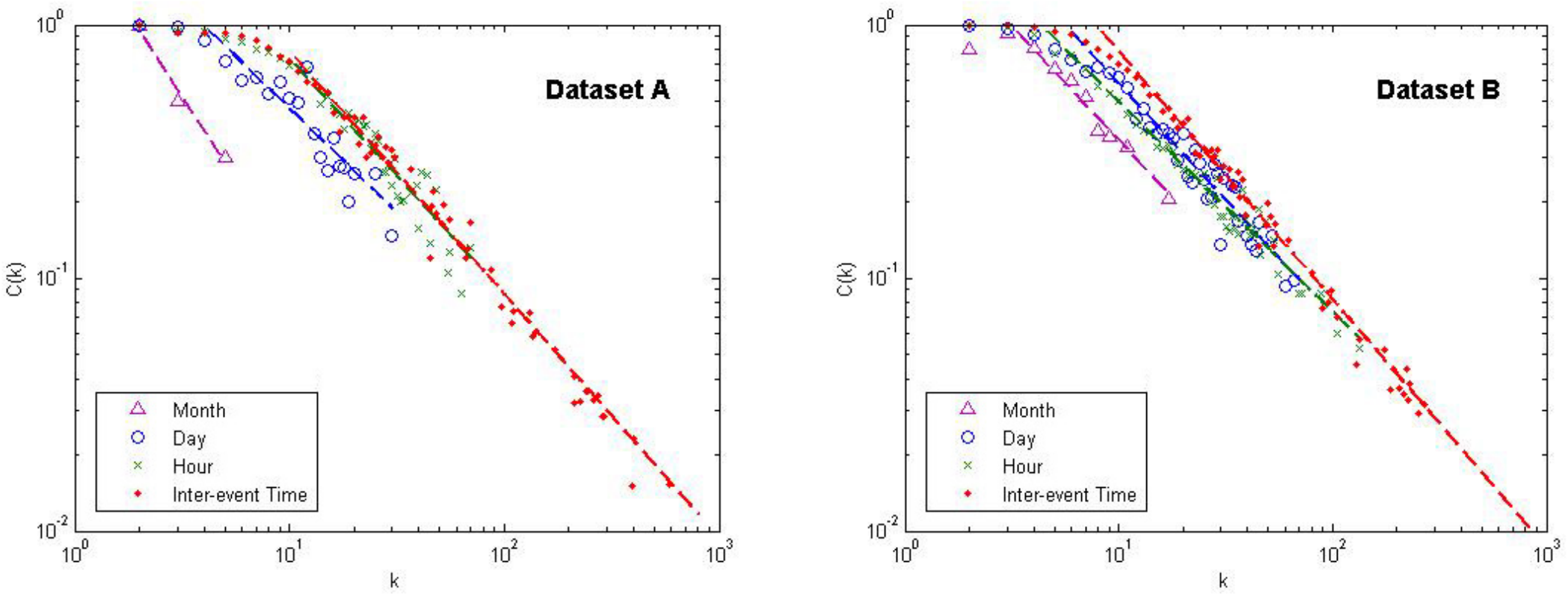

Figure 6. Plots of $\overline{C}(k) \sim k$ to investigate the hierarchical structure of networks.

## 5. Fractal and self-similarity analysis of visibility graphs

We have already proved that the time series of library loans have fractal character. Whether the visibility graphs also have that feature as well as self-similarity will be discussed in this section.

The relation between scale-free networks and self-similar networks has been studied recently [26-31], the fractal networks seem to exhibit scale-free property while scale-free networks are not always self-similar. First of all, it should be pointed out that fractality and self-similarity have the same meaning in the field of classic fractal geometry; however, they don't imply each other in the study of complex networks. That is to say, the fractal networks are always self-similar; whereas some networks are self-similar, that is, exhibit the scale-invariant degree distribution, and yet are not fractals. A typical example of such networks is the Internet. The fractal and self-similar nature of network can be revealed by the well-known box-counting method.

Here we use the node-covering box-counting method introduced by Song et al. [26] to calculate the minimum number of boxes needed to cover the whole network with box size $l_B$ where $l_B - 1$ is the maximum distance of all possible pairs of nodes in each box. In such rules the boxes should cover all the nodes of the network and each node belongs to only one box. If the minimum number $N_B$ of covering boxes scales with respect to box size $l_B$ as:

$$N_B(l_B) \sim l_B^{-d_B} \tag{9}$$

the network deem to be fractal , where $d_B$ denotes the fractal dimension or self-similar exponent.

Moreover, we can use the renormalization procedure, i.e. coarse graining through the box-counting method to verify a network is self-similar or not where very box is replaced with a single super-node and two boxes are then connected when there was at least one link between their constituent nodes. Repeat that process until the network is renormalized to one node. If the degree distribution of the network keeps invariant under the coarse graining by the boxes with different size along with successive applications of the coarse-graining transformation, the network is called self-similar.

According to the theory above, we covered the eight networks with different box size $l_B$ and obtained the number of boxes $N_B$. As shown in Fig.7, by fitting the curves of $N_B \sim l_B$, it can be clearly observed that $N_B$ decays with $l_B$ exponentially, which means all the visibility graphs don't have fractal structure. When the networks are renormalized, we also find that although the networks remain scale-free, the power-law exponents become smaller both when $l_B$ increases and under successive coarse-graining procedure. In consequence the networks are not self-similar, that is to say the networks as a whole can't be replaced by some part of it. Furthermore, it's not appropriate to use some sections of time series to reflect the global pattern of entire time series owning to that the parts and parts, as well as whole and parts are not analogous.

According to Song et al. [27], the network with hubs attraction will not be self-similar. On the contrary, the hub repulsion, i.e. the nodes with large degree tend to be connected with the ones have few links will bring the network with fractal feature and higher robustness. Whether the interaction is attraction or repulsion among the hubs of the network can be determined from the correlation between the degrees of different nodes. The degree correlation can be quantified by Pearson correlation coefficient $r$ [32] defined as:

$$r = \frac{M^{-1}\sum_i k_1 k_2 - \left[M^{-1}\sum_i \frac{1}{2}\left(k_1 + k_2\right)\right]^2}{M^{-1}\sum_i \frac{1}{2}\left(k_1^2 + k_2^2\right) - \left[M^{-1}\sum_i \frac{1}{2}\left(k_1 + k_2\right)\right]^2}$$

(10)

where $k_1$ and $k_2$ are the degrees of the nodes at the ends of edge $i$, and $M$ stands for the total number of edges in the network. There is no doubt that the degree correlations of adjacent nodes have strong effects on the structure of the network. The hubs in networks with positive correlation ($r > 0$) are attracted to each other with large probability while the interaction between hubs of a network with negative correlation ($r < 0$) are repulsive, and the networks are called assortative mixing such as social networks and disassortative mixing such as technical network respectively.

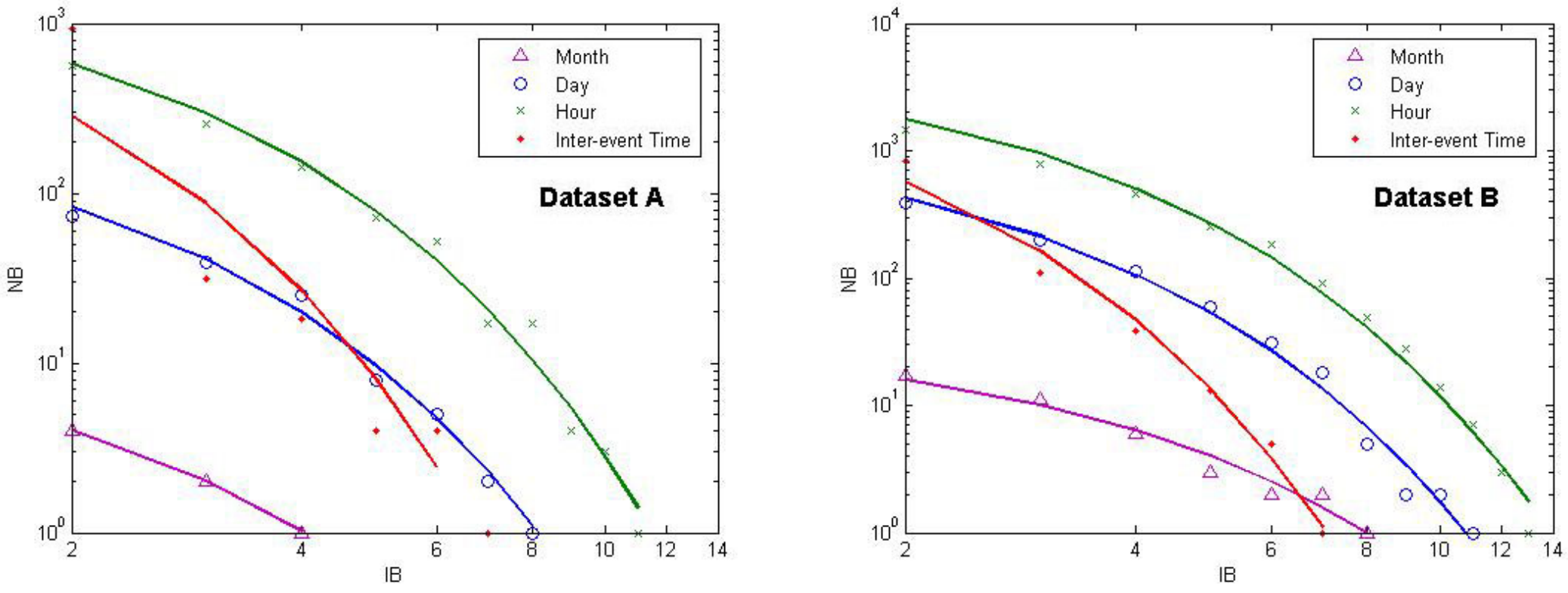

Figure 7. Plot of $N_B \sim l_B$ to investigate the fractal property of networks.

In our case, the values of $r$ are calculated following Eq. (10) and the results are added to Table 3. It's clear that among the 8 networks, 4 of them are hub-attractive and the others are hub-repulsive. That may conflict with the conclusion that all the networks are not fractal. Our work may raise some doubts about the conclusion that the fractal architecture of networks originates from the strong effective repulsion between hubs on all length scales. There may be some other unknown origin where the fractal structure stems from.

## 6. Summary and discussion

During its development, the studies on human dynamics give much emphasis on the temporal-spatial scaling law from individual, collective and group perspectives. Nevertheless, fractal characteristic has got some concern by many scholars. As early as 2000, Plerou et al. [33] found long-range correlation in stock price changes, that is to say, the trading activity tends to keep the same trend as past for a considerable time. Then Paraschiv-Ionescu et al. [34] investigated the fundamental pattern in human physical activity and found the fractal structure which may be disrupted by chronic pain. Recently, Rybski et al. [35] investigated the presence of temporal correlations in the individual activity of short-message communications, and showed that the dynamics of the more active members display clear long-term correlations. All the results imply that human behaviors exhibit fractal feature.

In our research, the amount of actions performed by people in a certain period of time is explored with different time scale from a brand new viewpoint. We testify the fractal property in human behavior dynamics from two different perspectives: Firstly, we use Rescaled Range Analysis to calculate the Hurst exponents and the length of non-periodic cycles, and find that for all the time series constructed from library loans the long-range correlation exponent $H > 0.5$, therefore the number of actions performed by people is fractal time series with long-term correlation. Secondly, the complex networks converted from time series with visibility algorithm exhibit scale-free property, small-world effect and hierarchical structure which confirm that the original time series are fractal within which data points are intensively connected with each other.

All the results suggest that time series of human behaviors do not follow random walk but own some intrinsic regularity within it. Memory plays a significant role in human activity to make the future keep the same trend as past to a certain extent. After visualized, the data points in time series with larger library loans than usual have been converted to hub-nodes in complex networks, and the interaction between them has significant influence on the property of both networks and time series. Regarding that the extreme points correspond to some special day in real life, the conclusion of our work may be helpful to manage and forecast the repetitious behaviors on collective level such as communication traffic control and computer network server design.

## Acknowledgments

This paper is supported by the National Natural Science Foundation of China (Grant No.70871082) and the Foundation of Shanghai Leading Academic Discipline Project (Grant No.S30504). We also thank Wang Li-na for helpful discussion.